\documentstyle[11pt]{article}
\begin{document}
\thispagestyle{empty}
%\pagenumbering{arabic}
%\large
\baselineskip=20pt
%---------------------------------------------------------------------------

\hfill{
\begin{tabular}{l}
DSF$-$98/5 \\
INFN$-$NA$-$IV$-$98/5\\
physics/9803040
\end{tabular}}

\bigskip\bigskip

\begin{center}
\begin{huge}
{\bf Radiation Induced Fermion Resonance }
\end{huge}
\end{center}

\vspace{2cm}

\centerline{\bf{S. Esposito}$^{1}$, \bf{M. W. Evans}$^{2}$, {\bf and  E. 
Recami}$^{345}$}

\medskip
                                 
\centerline{$^{1}${\it I.N.F.N., Sezione di Napoli},{\it 80125 Naples, Italy}}
\centerline{$^{2}$ {\it JRF 1975},{\it Wolfson College, Oxford, Great Britain}}
\centerline{$^{3}${\it Facolt\`{a} di Ingegneria}, {\it Universit\`{a} 
Statale di Bergamo}, {\it 24044 Dalmine (Bergamo) Italy}}
\centerline{$^{4}${\it INFN Sezione di Milano, Milan, Italy}}
\centerline{$^{5}${\it C.C.S. and D.M.O./FEEC}, {\it UNICAMP, 
Campinas, Sao Paolo, Brazil}}

\vspace{1truecm}

\section*{\bf{Abstract}}

The Dirac equation is solved for two novel terms which describe the 
interaction energy between
the half integral spin of a fermion and the classical, circularly polarized, 
electromagnetic field. A
simple experiment is suggested to test the new terms and the existence of 
radiation induced
fermion resonance.

\smallskip

\centerline{-------------------------------- }

\bigskip

Recently, Warren {\it et al.} [1,2]  have made the first attempt to detect 
radiation induced fermion
resonance due to irradiation by a circularly polarized electromagnetic field 
[3-6].  In this Letter
the Dirac equation is solved for one fermion in a classical electromagnetic 
field.  Two new terms
are inferred which show the theoretical existence of radiation induced 
fermion resonance, and the
experimental conditions under which this phenomenon can be detected are 
defined for an electron
beam.

The demonstration is based on the standard Dirac Hamiltonian operator 
(Gaussian units),

\smallskip

$$
{\rm H} \: = \: c {\vec{\alpha}} \: \cdot \: 
\biggl( {\vec{p}} \, - \, \frac {\rm e}{\rm c} 
{\vec{A}} 
\biggr)
\: + \: \beta 
m c^{2} \: + \: e V \, , \eqno {\bf(1)}
$$

\noindent describing a 
fermion (e.g. an electron) of mass m and charge e interacting with a
classical electromagnetic field with vector potential 

\smallskip

$$
A^{\mu} \: = \: \biggl ( V, \, \, \vec{ A}  \biggr ) \, \, .
\eqno {\bf(2)}
$$

\noindent through the eigenvalue equation,

\smallskip

$$
{\rm H} \psi \: = \: \rm{E} \psi \, ,
\eqno {\bf(3)}
$$

\noindent Here 
$\vec{\alpha}$  and $\beta$ are the usual Dirac matrices and $\psi$ the 
four-component Dirac spinor. The rest energy of the fermion is $mc^2 $   
and its three momentum is
$\vec{p}$, as usual.

In the usual non-relativistic approximation the calculation proceeds by 
setting up Eq. (3) for the
proper dominant wavefunction, $\phi $ ,

\smallskip

$$
{\rm H}^{\prime} \phi \: = \: \rm{E} \phi \, \, ,
\eqno {\bf(4)}
$$

\noindent Writing the 4-component Dirac spinor 

\smallskip

$$
\psi \: = \: \left( \begin{array}{c}
\psi_A \\
\psi_B  \end{array} \right)
\eqno {\bf(5)}
$$

\noindent where $\psi_A$ and $\psi_B$ are respectively the ``large''
and ``small'' component satisfying the condition (at second order in
$v/c$)

\smallskip

$$
\int \left( \psi^{\dagger}_A \psi_A \, + \, \psi^{\dagger}_B \psi_B 
\right) d^3 x \; \simeq \; \int \psi^{\dagger}_A 
\left( 
1 \, + \, \frac {\rm \vec{\pi}^{2} } {\rm 8 \,  m^2 \, c^2 } 
\right) \psi_A \, d^3 x \; \simeq \; 1
\eqno {\bf(6)}
$$

\noindent (here $\vec{\pi} \, = \, \vec{p} \, - \, \frac{e}{c} \vec{A}$),
the dominant two-component wavefunction $\phi$ in the non relativistic limit
is given by (again at second order in $v/c$)

\smallskip

$$
\phi \; \simeq \; \left( 
1 \, + \, \frac {\rm \vec{\pi}^{2} } {\rm 8 \,  m^2 \, c^2 } 
\right) \psi_A
\eqno {\bf(7)}
$$

By standard methods of
solution we find that the Hamiltonian in Eq. (4) is made up of six 
terms as follows,

\smallskip

$$
{\rm H} \: = \: \frac {\vec{\pi}^2} {2m} 
\, + \, e V \, - \, \frac {\rm e \hbar} { \rm 2mc} \vec{\sigma} \, 
\cdot \, \vec{B} \; -
\eqno {\bf( H1)}
$$

\smallskip

$$
- \, \frac{\vec{\pi}^4} { \rm 8 m^2 c^2 } \; -
\eqno {\bf( H2)}
$$

\smallskip

$$
- \, \frac { \rm e \hbar^2 } { \rm 8 m^2 c^2 } 
\biggl (
\vec{{\bf \bigtriangledown}} \, \cdot \, { \vec{E} } \, + \, 
\frac {\rm 1} {\rm c} 
\frac {\rm \partial} { \rm \partial t} \biggl ( \vec{{\bf \bigtriangledown}} 
\, \cdot \, {\vec{A}} \biggr)
\biggr ) \; -
\eqno {\bf( H3)}
$$

\smallskip

$$
- \, \frac {\rm e \hbar} {\rm 4 m^2 c^2} \vec{\sigma} \, \cdot \, 
\vec{\epsilon } \: \wedge \:
{ \vec{p} } \; + 
\eqno {\bf( H4)}
$$

\smallskip

$$
+ \, \frac {\rm e^2 \hbar} {\rm 4 m^2 c^3 } \vec{\sigma} \, \cdot \, 
{\vec{E}}
\: \wedge \: {\vec{A}} \; -
\eqno {\bf( H5)}
$$

\smallskip

$$
- \, \frac {\rm e^2 \hbar } { \rm 4 m^2 c^4 } \vec{\sigma} \, \cdot \, 
\biggl (
{\vec{ A} } \: \wedge \: \frac {\rm \partial {\vec{A}} } 
{\rm \partial t } \biggr ) 
\, \, .
\eqno {\bf( H6)}
$$

\noindent Here   ${\vec{B}} \: = \: \vec{{\bf \bigtriangledown}} \: \wedge \: 
{\vec{ A} }$ is the magnetic field  ; and $ \vec{\epsilon} \: = \:
{\vec{ E}} \, + \, \frac {\rm 1} {\rm c} \,
 \frac {\rm \partial {\vec{A}} } {\rm 
\partial t}   $  ; where  
${\vec{ E}} \: = \: - \, \vec{{\bf  \bigtriangledown}} V \, - \, 
\frac {\rm 1} {\rm c} \,
 \frac {\rm \partial {\vec{A}} } {\rm 
\partial t} $ is the electric field.

\smallskip

These terms can be interpreted as follows: (H1) is the well-known
Schr\"{o}dinger-Pauli Hamiltonian; (H2)
is the relativistic correction to the kinetic energy; (H3) is the Darwin term; 
(H4) gives the spin-orbit
coupling term. These four terms are well known corrections to order 
$ (v / c)^2$   of the Schr\"{o}dinger equation (in the electrostatic 
case). Terms (H5) and (H6) are novel, 
and represent the coupling between the half
integral fermion spin and the circularly polarized classical electromagnetic 
field. They give rise to
radiation induced fermion resonance, and we will focus on them.

In S.I. units, terms (H5) and (H6) both give rise to the interaction 
eigenenergy,

\smallskip

$$
\rm{H} \: = \: \frac {\rm \mu_{0} \hbar} {\rm 4 c } 
\biggl ( \frac {\rm e} {\rm m} 
\biggr )^2
\frac {\rm I} {\rm \omega} \, \vec{\sigma} \, \cdot \, { \vec{ k}}
\eqno {\bf( 8)}
$$

\noindent where $I$  is the beam power density $ (W m^{-2} )$ and $\omega$  
its angular frequency ($\vec{k}$ is the z-axis versor). 
Here $\mu_{0}$  is the vacuum permeability. In deriving Eq. (6) 
it has been assumed
that ${\vec{ E}}$ and ${\vec{ A}}$ are plane waves in the vacuum and that

\smallskip

$$
I \: = \: \frac {\rm c } {\rm \mu_0} \, B^{(0)2} \, \, ,
\eqno {\bf( 9)}
$$

\noindent where $B^{(0)}$   is the scalar amplitude of ${\vec{  B}} \: = \: 
\vec{{\bf \bigtriangledown}} \: \wedge \: {\vec{ A}}$.  
The resonance frequency from Eq. (8) is calculated from spinor states as 
usual and is

\smallskip

$$
\omega_{res} \: = \: \frac {\rm \mu_0 } {\rm 2 c } \, \biggl ( \frac {\rm e}
{\rm m } \biggr )^2 \frac {\rm I} {\rm \omega} \: = \: 6 \: \cdot \: 4832 \: 
\times \: 10^7
\frac {\rm I} {\rm \omega}\, \, ,
\eqno {\bf( 10)}
$$

\noindent for the electron.

For example, if $I$  is $100 $ watts per square centimeter ($10^6 \:  
W m^{-2} $  ), and $\omega $ 
is tuned to 1.2815 MHz, $\omega_{res}$     occurs at the same frequency. 
This means that
the resonance absorption can be detected as a decrease in the r.f. output 
applied to an electron
beam. This simple experiment tests the existence of terms (H5)and (H6) 
provided that the r.f.
output is accurately circularly polarized. There is no effect expected in 
linear polarization. 

The clear importance of a positive result to this experiment would be that 
site specific shifts [1,2]
of easily a megahertz or more could be induced by irradiation in an ESR 
spectrum, giving a new
analytical technique. The prediction in Eq. (8) can be tested to probably one 
part in $10^9 $  if the
resonance frequency can be resolved to a hertz or less. This gives a 
very severe test in
fundamental physics of the semi-classical Dirac equation; and also of 
its presumably more
accurate counterparts in quantum electrodynamics and unified field theory. 

Similarly, site specific shifts in the range of perhaps one to a thousand 
hertz can be induced in an
NMR spectrum [1,2] by pulsing with circularly polarized r.f. radiation. An 
NMR spectrometer has
sub-hertzian resolution, so these site specific shifts should, in theory, 
lead to a new analytical
technique as was the purpose of Refs. [1] and [2]. 

\section*{\bf {Acknowledgments}}

Internet discussions over the past two years are gratefully acknowledged 
with many leading
specialists. The Indian Statistical Institute and Alpha Foundation are 
thanked for positions,
honoris causa, for MWE. Funding of this work by INFN is gratefully 
acknowledged.
   
\section*{\bf {References}}
\begin{description}
\item{[1]}{ W. S. Warren, S. Mayr, D. Goswami, and A. P. West, Jr., 
{\it Science} $\bf 255$, 1683 (1992).}
\item{[2]}{ {\it ibid.}, $\bf 259$, 836 (1993).}
\item{[3]}{ M. W. Evans, {\it J. Phys. Chem.} $\bf 95$, 2256 (1991)}.
\item{[4]}{ D. Goswami, Ph. D. Thesis, Princeton, 1994.}
\item{[5]}{ M. W. Evans and S. Kielich, (eds.), {\it Modern Nonlinear Optics}, 
Vol. 85(2) of {\it
Advances in Chemical Physics}, I. Prigogine and S. A. Rice, eds., 
(Wiley Interscience, New York,
1997, paperback printing).}
\item{[6]} {P. A. M. Dirac, {\it Quantum Mechanics}, 4th edn.  
(Oxford University Press,
Oxford, 1974).}
\end{description}
\end{document}